\documentstyle[12pt]{article}
\advance \textheight by \topskip
\begin{document}
\begin{titlepage}
\hbox to \hsize{{\hfil IHEP 97-31}}
\hbox to \hsize{{\hfil hep-ph/9705378}}
\hbox to \hsize{{\hfil May, 1997}}
\vfill
\large \bf
\begin{center}
Different behaviour of the spin structure functions
 $g_1(x)$ and $h_1(x)$ at $x\to 0$.
\end{center}
\vskip 1cm
\normalsize
\begin{center}
{\bf S. M. Troshin\footnote{e-mail: troshin@mx.ihep.su } and N. E. Tyurin\footnote{e-mail: tyurin@mx.ihep.su }}\\[1ex]
{\small  \it Institute for High Energy Physics,\\
Protvino, Moscow Region, 142284 Russia}
\end{center}
\vskip 1.5cm
\begin{abstract}
We consider low-$x$ behaviour of the spin structure functions
$g_1(x)$ and $h_1(x)$ in the unitarized chiral quark model which
combines ideas on the constituent quark structure of hadrons with
a geometrical scattering picture and unitarity.  A nondiffractive
singular low-$x$  dependence of $g^p_1(x)$ and $g_1^n(x)$ indicated
by the recent SMC
experimental data is described. A diffractive type
smooth behaviour of $h_1(x)$  is  predicted at small $x$.
The expectations for the double-spin
asymmetries in the low-mass Drell-Yan production at RHIC in the
central region are discussed alongside.\\ [2ex]
\end{abstract}
\vfill
\end{titlepage}
\section*{Introduction}
Low-$x$ behaviour of the spin structure functions
$g_1(x)$ and $h_1(x)$ is important for understanding
 a nucleon spin structure. Experimental evaluation
of the first moment of $g_1$ (and
 the total hadron helicity carried
by  quarks) is sensitive to  extrapolation of $g_1(x)$ to $x=0$.
This extrapolation is a nontrivial issue
 in the light of the recent CERN SMC
 \cite{smc} and
SLAC E154 \cite{e154} data  indicating a rising
 behaviour of $g_1(x)$ at small $x$.

Usually a smooth Regge extrapolation (corresponding to the contribution
of the Regge pole of unnatural parity)
$g_1\sim x^{-\alpha_{a_1}}$ ($-0.5<\alpha_{a_1}<0)$
being used for extrapolation of the data to $x=0$.
It is evident that the CERN SMC and SLAC E154 experiments
give a little support for such smooth behaviour \cite{ell,ter}
and rather suggest a singular dependence on $x$ at $x\to 0$.

 A non-Regge behaviour of $g_1(x)$ at small $x$
and a connection  of such behaviour with a diffractive contribution is
under discussion \cite{clo} since the first EMC data have
been published \cite{emc}. There were used various explicit functional
dependencies  based either on the
perturbative QCD considerations \cite{ell,rys}
or the nonperturbative approaches \cite{clo1,land}.
These parametrizations can successfully
describe increasing behaviour of $g_1^p(x)$ at small values of $x$.
Upper bounds for the low-$x$ behaviour of the spin structure
functions were also obtained \cite{clo1,tro} using the general
properties of the scattering matrix and some heuristic
considerations.

However, the above approaches relate this increasing behaviour with
the diffractive contribution to $g_1(x)$ at
 small $x$ which seem
does not find confirmation in the recent experiments. In particular,
 the SMC data demonstrate the following equality\footnote{
Preliminary SMC results \cite{magn} based on the data obtained in 1996
indicate a possibility for a
 more complicated behaviour of the spin structure function $g_1^p(x)$.}
  in the region of
$0.003\leq x \leq 0.1$:
\begin{equation}
g_1^p(x)=-g_1^n(x).\label{pmn}
\end{equation}
It might happen that the leading diffractive contribution providing
the same sign
input to the proton and neutron structure functions would not be able
to serve as a sole explanation of the observed experimental
regularities at $x\to 0$, in particular, Eq. (\ref{pmn}).
Combining the existing facts one could assume importance of
 a nondiffractive
behaviour of $g_1(x)$ at small $x$.

The  essential point  is that the space-time structure of
the scattering
at small values of $x$ involves the large distances
$l\sim 1/Mx$ on the light--cone \cite{pas} and the
region $x\sim 0$ is therefore sensitive to the nonperturbative dynamics.
It leads to necessity
of applying the nonperturbative model approaches.

One should note that the general principles such as unitarity
and analyticity are  useful  and  provide
some constraints.
  In particular, unitarity provides the following upper
 bounds for $g_1(x)$ and $h_1(x)$ \cite{tro}:
\begin{equation}
g_1(x)\leq\frac{1}{x}\ln(1/x)\quad\mbox{and}\quad h_1(x)\leq
\frac{1}{x}\ln(1/x).\label{upb}
\end{equation}

The experimental data for
$\Delta\sigma _L(s)$
and
$\Delta\sigma _T(s)$
also could be a useful source of information
on the low-$x$ behaviour of the spin structure functions. Unfortunately,
only the low energy data are available at the moment \cite{krs}. Were the
experimentally observed decreasing behaviour of
$\Delta\sigma _L(s)$
and
$\Delta\sigma _T(s)$
 also valid at high energies, it would be possible to conclude
that:
\begin{equation}
xg_1(x)\to 0\quad\mbox{and}\quad xh_1(x)\to 0\label{zer}
\end{equation}
at $x\to 0$.

In this paper we show that the non-Regge, nondiffractive behaviour
of $g_1(x)$ can be  described in the unitarized chiral
quark model \cite{us} which
combines ideas on the constituent quark structure of hadrons with
a geometrical scattering picture and unitarity. Different functional
dependence is predicted for the structure function $h_1(x)$ at small $x$
which contrary to $g_1$ has a diffractive origin and satisfies to
the equality
$h_1^p(x)=h_1^n(x)$ at small $x$.
For that purpose we consider the corresponding quark
 spin densities $\Delta q(x)$ and $\delta q(x)$.
We discuss also some predictions which might
be interesting
for the forthcoming RHIC spin experiments.
Namely,   the quark spin densities
$\Delta q(x)$ and $\delta q(x)$  at small $x$
determine  behaviour of the spin asymmetries in
 hadron-hadron interactions, in particular, double--spin asymmetries
$A_{LL}$ and $A_{TT}$ which are to be measured at RHIC
in Drell-Yan processes with low-mass  lepton pairs.

\section{Outline of the model}
To obtain the explicit forms for the quark spin densities
 $\Delta q(x)$ and $\delta q(x)$ it is convenient to use the relations
 between these functions and  discontinuities in the helicity
amplitudes of the forward antiquark--hadron scattering \cite{jaffe}
which are based on the dominance of the ``handbag'' diagrams in
deep--inelastic processes:
\begin{eqnarray}
q(x) & = & \frac{1}{2}\mbox{Im}[F_1(s,t)+F_3(s,t)]|_{t=0},\nonumber \\
\Delta q(x) & = & \frac{1}{2}\mbox{Im}[F_3(s,t)-F_1(s,t)]|_{t=0},\nonumber \\
\delta q(x) & = & \frac{1}{2}\mbox{Im} F_2(s,t)|_{t=0} \label{h1},
\end{eqnarray}
where $s\simeq Q^2/x$ and $F_i$ are the helicity amplitudes for the
elastic quark--hadron scattering in the notations for
the nucleon--nucleon scattering, i.e.
\[
F_1\equiv F_{1/2,1/2,1/2,1/2},
\, F_2\equiv F_{1/2,1/2,-1/2,-1/2}, \,
F_3\equiv F_{1/2,-1/2,1/2,-1/2},
\, F_4\equiv F_{1/2,-1/2,-1/2,1/2}
\]
and
\[ F_5\equiv F_{1/2,1/2,1/2,-1/2}.
\]

We consider quark as a structured hadronlike object since at small
$x$ the photon converts to a quark pair at  large distances before
it interacts with the hadron. At large distances perturbative QCD
vacuum  undergoes  transition into a nonperturbative one with
formation of the  quark condensate. Appearance of the condensate
means the spontaneous   chiral symmetry breaking and the current quark
transforms into a massive quasiparticle state -- a constituent quark.
Constituent quark is embedded into the nonperturbative vacuum (condensate)  and therefore we 
treat it similar to a hadron.
Arguments
in favour of such picture can be found in \cite{pas,us,deld}.
The quark--hadron scattering at small $x$ can be considered
similar to the hadron--hadron scattering.

Unitary representation for the helicity amplitudes follows
from their relations \cite{abk} to the $U$--matrix.
In the impact parameter representation:
\begin{equation}
 F_{\Lambda_1,\lambda_1,\Lambda_2,\lambda_2}(s,b)  =
U_{\Lambda_1,\lambda_1,\Lambda_2,\lambda_2}(s,b)+
i\rho (s)\sum_{\mu,\nu}
U_{\Lambda_1,\lambda_1,\mu,\nu}(s,b)
F_{\mu,\nu,\Lambda_2,\lambda_2}(s,b),\label{heq}
\end{equation}
where $\lambda_i$ and $\Lambda_i$ are the quark and hadron helicities,
respectively, and $b$ is the impact parameter of quark-hadron scattering.
The kinematical factor $\rho(s)$ is unity at high energies.
Explicit solution of  Eqs. (\ref{heq}) has a rather
complicated form,
however, in the approximation when the helicity-flip functions
are  less than the helicity nonflip ones
we can get simple expressions
\begin{eqnarray}
F_{1,3}(s,b) & = &{U_{1,3}(s,b)}/
{[1-iU_{1,3}(s,b)]},\label{f13}\\
F_2(s,b) & = & {U_2(s,b)}/{[1-iU_1(s,b)]^2}.\label{f2}
\end{eqnarray}
Unitarity requires  Im$U_{1,3}(s,b)\geq 0$.
The functions $F_i(s,t)$ are the corresponding Fourier--Bessel
transforms of the functions $F_i(s,b)$:
\begin{equation}
F_{i}(s,t)=\frac{s}{\pi^2}\int_0^\infty bdb F_{i}(s,b)
J_0(b\sqrt{-t}),\label{imp}
\end{equation}
where $i=1,2,3$.

We consider now the main points of the model \cite{us} which
allows to get explicit form for the $U$--matrix.
A hadron   is consisting of the
constituent quarks located at the central part  
 embedded into a nonperturbative vacuum (quark condensate).
Justification for this picture is the effective QCD approach and, 
in particular,
the Nambu--Jona-Lasinio  model  with the six--fermion
$U(1)_A$--breaking term.
The constituent quark
masses can be expressed in terms of the quark condensates.
Such quark appears  as a quasiparticle,
i.e. as a current quark
and the surrounding  cloud of the quark--antiquark pairs of the
different flavours.
Quantum numbers of the
 respective
constituent quarks are the same as the quantum
numbers of current quarks due
to  conservation of the corresponding currents in QCD.
The only exception
is the flavour--singlet, axial--vector current.
Quark radii are determined by the sizes of  the
respective clouds.  We take  the strong interaction radius
 of  quark  $Q$  as its Compton wavelength:
 $r_Q=\xi /m_Q$, where
 constant $\xi$ is universal for the different  flavours. Quark
 formfactor $F_Q(q)$ is taken in the dipole form and
 the corresponding quark matter
distribution $d_Q(b)$ has the form \cite{us}
 $d_Q(b)\propto \exp(-{m_Qb}/{\xi})$.
Spin of the constituent quark $J_{U}$  in this approach is given
by the  following sum \begin{equation}
J_{U}=1/2  =  J_{u_v}+J_{\{\bar q q\}}+\langle L_{\{\bar qq\}}\rangle              =  1/2+J_{\{\bar q q\}}+\langle L_{\{\bar qq\}}\rangle.
\label{bal}
\end{equation}

 The  current quark helicity is reduced by the
negative contribution of the helicities of quarks
 from the cloud in the structure of constituent quark.
So we can
say that significant part of the spin of constituent quark is
associated with the orbital angular momentum of quarks inside this
the constituent quark, i.e. the cloud quarks  rotate coherently inside
 their constituent quark.
Thus,  we assume the standard $SU(6)$ spin structure
of a nucleon
consisting of the three constituent quarks (embedded into the
condensate), i.e. all the nucleon spin
is composed from the spins of the constituent quarks.
Constituent quarks however have
a complex internal spatial  and spin structure.

   The picture of hadron consisting of constituent quarks and
 surrounding condensate implies that overlapping and interaction of
 peripheral clouds occur at the first stage of hadron interaction.
 Condensates in the overlapping region are
  excited  and  as  a result the quasiparticles , i.  e.
massive  quarks appear in the overlapping region.
It should be noted that
the condensate excitations are massive quarks,
since the vacuum is nonperturbative one and there is no overlap between
 the physical (nonperturbative) and bare (perturbative) vacuum.
 The  part of
 hadron  energy  carried  by  the  outer clouds of condensates being
released  in  the overlapping region, goes to the generation of
massive quarks.  Number of such quarks fluctuates.  Their average
number  in the framework of the geometrical
 picture can be estimated as follows:  \begin{equation} N(s,b)
\propto N(s)\cdot D^{h_1}_c\otimes D^{h_2}_c. \label{4} \end{equation}
  The function $D^{h_i}_c$
describes condensate distribution inside hadron $h_i$; $b$ is the
impact parameter of colliding hadrons $h_1$ and $h_2$.
For the
function $N(s)$ we  use the maximal possible value
$N(s)\propto\sqrt{s}$.
Thus, $N(s,b)$ of massive virtual quarks appear in the
overlapping region and  they generate  effective field.
There is another part of the effective field associated with
the selfconsistent field of the constituent quarks.

Constituent quarks  are supposed to scatter in a
quasi-independent way by this effective two--component field.
We can write down  explicit form for the
function $U(s,b)$ for quark-hadron scattering.
For simplicity we consider helicity nonflip part of $U$--matrix
as a pure imaginary one.
In accordance with the
quasi-independence of valence quarks we represent the input
dynamical  quantity in  the  form  of the  product
\cite{us}:
\begin{equation}
U(s,b)\,=\, \prod^{n+1}_{Q=1}\,f_Q(s,b)\label{fac}
\end{equation}
in the impact parameter representation. The factors $f_Q(s,b)$
in Eq. \ref{fac} correspond to the individual quark amplitudes.
The function $f_Q(s,b)$ describes the scattering of a single
valence quark  in the effective field generated by  virtual
and  valence  quarks  as  it  was discussed above.
Combining the above arguments about constituent quark
scattering the function $U(s,b)$
 can be written  as follows \cite{us}
\begin{equation}
U(s,b)=i\left[ \gamma n+\frac{a\sqrt{s}}{\langle m_Q\rangle}\right]^{n+1}
\exp(- Mb/\xi).\label{uf}
\end{equation}
In Eq. (\ref{uf})
$n$ is the  number of the constituent quarks in the
hadron, $M=\sum_{Q=1}^{n+1} m_Q$, and $b$ is the
impact parameter of the colliding quark and hadron \cite{us}.
  $a$ is a  universal constant for different hadrons while the
constant $\gamma$ depends on the types of the colliding hadrons.

Mechanism of quark helicity flip in this picture is associated with
the constituent quark interaction with the quark generated under
interaction of the condensates \cite{abk}.
 Quark exchange process between the valence quark and an
appropriate quark with relevant orientation
of its spin and the same flavour will provide the necessary helicity
flip transition, i.e. $Q_+\rightarrow Q_-$.

Of course, such processes are relatively suppressed in comparison
with the quark scattering preserving
helicity  and therefore do not  contribute into to the
spin-averaged observables at the leading order. However, measuring of the transverse
 spin asymmetries is sensitive to the subleading contribution and serves therefore as a filter of the quark exchange
processes  $Q_+\rightarrow Q_-$.
This transition occurs when the valence   quark
knocks out  a quark with the opposite  helicity
and the same  flavour.  This  interaction    has
energy suppression by the factor $1/N(s)$ and it
results in the following relation between
the functions $U_1(s,b)$ and $U_2(s,b)$:
\begin{equation}
\frac{U_2(s,b)}{U_1(s,b)}\sim
\frac{\langle m_Q\rangle ^2}{s}\exp[2(\alpha-1)\langle m_Q\rangle b/\xi].
 \label{u}
\end{equation}
The value of the parameter $\alpha>1$ reflects the more
central character
of the quark helicity flip mechanism.

Now we return to  considerations of the helicity nonflip functions
$U_1$ and $U_3$. These functions differ in the helicities of the initial
and the final states but both describe helicity nonflip scatterings.
The second term in the square brackets of Eq. (\ref{uf}) results from
the quark interaction with the component of the effective field which in
its turn arises from the interaction of the condensates.
Since the hadron spin is
composed from the spins of the constituent quarks, this part of
interaction does not depend on the quark spin orientation and the second
term in the square bracket in Eq. (\ref{uf}) should be the same for the
functions $U_1$ and $U_3$ due to the parity conservation. This argument
does not work for the first term
 which follows from the quark
interaction  with selfconsistent field of the valence quarks.
 Thus, we should
suppose that this part of the effective field
depends on the relative spin orientations of the valence quarks
and consequently the constant $\gamma$, in fact,
 has different values in the expressions
for the functions
$U_1$ and $U_3$.
It is flavour dependent also.
Thus, instead of Eq. (\ref{uf}) we have
\begin{equation}
U_{1,3}(s,b)=i\left[\gamma_{1,3}n+\frac{a\sqrt{s}}{\langle m_Q\rangle}
\right]^{n+1}
\exp(-Mb/\xi).\label{u13}
\end{equation}
Then Eq. (\ref{h1}) allows to obtain the quark densities
$\Delta q(x)$ and
 $\delta q(x)$ at small $x$.

\section{Results and discussion}

Calculating $F_{1,2,3}(s,t)|_{t=0}$ at high values of $s$ we get
$q(x)$, $\Delta q(x)$ and $\delta q(x)$ at small $x \simeq Q^2/s$, i.e.
\begin{equation}
q(x)\sim \frac{1}{x}\ln^2(1/x)\label{df1},
\end{equation}
\begin{equation}
\Delta q(x)\sim \frac{1}{\sqrt{x}}\ln(1/x)\label{dg1}
\end{equation}
and
\begin{equation}
\delta q(x)\sim x^{\frac{\alpha-1}{n+1}}\ln(1/x).\label{dh1}
\end{equation}
The behaviour of
 $ q(x)$ (and $F_1(x)$) and
 $\delta q(x)$ (and  $h_1(x)$) is
 determined by the leading terms in $U_{1,3}(s,b)$ and $U_2(s,b)$
 and therefore
the small-$x$ dependence of $F_1(x)$ and $h_1(x)$
will be the universal for the proton and neutron, i.e.:
\begin{equation}
F^p_1(x)=F_1^n(x)\sim
\frac{1}{x} \ln^2(1/x).\label{f1m}
\end{equation}
and
\begin{equation}
h^p_1(x)=h_1^n(x)\sim
 x^{\frac{\alpha-1}{n+1}}\ln(1/x).\label{h1m}
\end{equation}
These functions results from the leading terms and have
a features of a diffractive origin.
It is also seen that $h_1(x)$ has a smooth behaviour at $x\to 0$, i.e.
$h_1(x)\to 0$ in this limit ($\alpha>1$).

The behaviour of $\Delta q(x)$ and correspondingly $g_1(x)$ is
 determined  by the difference
$U_3(s,b)-U_1(s,b)$ or the subleading terms in $U_{1,3}$
and therefore
the small-$x$ dependence of $g_1(x)$ in this model shows
different constant factors for the proton and neutron, i.e.:
\begin{equation}
g_1^{p,n}(x)\simeq \frac{C^{p,n}}{\sqrt{x}}\ln(1/x).\label{gpn}
\end{equation}
Contrary to $h_1$ the spin structure function $g_1$ has a singular
behaviour at $x\to 0$.

The magnitude and sign of the constants $C^{p,n}$ do not follow from
the model. Fit to the SMC data provides  a good agreement of
Eqs. (\ref{gpn}) with experiment at small $x$
($0<x<0.1$) (cf. Fig. 1,2) and also leads to $C^p=-C^n=0.021$.

The first moment
 \begin{equation}
\Gamma_1=\int_0^1g_1(x)dx
\end{equation}
exists. To evaluate $\Gamma_1^p$ we use the value of
\begin{equation}
I(0.1,1)=\int_{0.1}^1g_1^p(x)dx=0.092,
\end{equation}
obtained with the standard parameterization of the
data at medium and large values of $x$ \cite{clo1}.
Eq. (\ref{gpn}) then gives
\begin{equation}
I(0,0.1)=\int_{0}^{0.1}g_1^p(x)dx=0.057
\end{equation}
and for the first moment
$\Gamma_1^p$
we obtain
\begin{equation}
\Gamma_1^p=\int_0^1g_1^p(x)dx=0.149
\end{equation}
The above simple estimation of
$\Gamma_1^p$ alongside with the known values of $g_A=1.257$ and $3F-D=0.579$ provide the following
approximate values for the quark spin contributions:
\begin{equation}
\Delta\Sigma  \simeq  0.25,\quad
\Delta u  \simeq  0.81,\quad
\Delta d  \simeq  -0.45,\quad
\Delta s  \simeq  -0.11, \label{sp}
\end{equation}
which are in agreement with the results obtained in the comprehensive
analysis with account for the QCD evolution and higher twist contributions
\cite{smc,ell,rid}. Eq. (\ref{sp}) demonstrates that the singular behaviour
of $g_1^p(x)$ in the form Eq. (\ref{gpn}) does not lead to significant
deviations from the results of the experimental analysis \cite{smc}
where the smooth extrapolation of the data to $x=0$ is used.
Nevertheless,  the available SMC data in the region of
$x\sim 10^{-3}$ provide a reasonable constraint for the functional form
of the spin structure function $g_1$ at small $x$.  We would
like to note that the functional dependence of
$g_1(x)\sim\frac{1}{\sqrt{x}}\ln(1/x)$
is in agreement with the E154 data as well.

It is important to note that
Eqs. (\ref{dg1}) and (\ref{dh1})
demonstrate
that the usual assumption
$\Delta q(x)\simeq\delta q(x)$
is not valid
 at small $x$.
The explicit forms
Eqs. (\ref{dg1}) and (\ref{dh1})
lead to the conclusion that the longitudinal and transverse
double--spin asymmetries for the Drell-Yan processes at $x_F\simeq 0$ are
 small,
i.e.
\begin{equation}
A_{LL}^{l\bar l}\simeq 0\quad
\mbox{and}\quad
A_{TT}^{l\bar l}\simeq 0
\end{equation}
when invariant mass of the lepton pair
$M^2_{l\bar l}\ll s$.
It is just this kinematical region which is sensitive to low-$x$
behaviour of the spin densities $\Delta q(x)$ and $\delta q(x)$
since:
\begin{equation}
A_{LL}^{l\bar l}=-\frac{\sum_ie_i^2[\Delta q_i(x_1)\Delta\bar q(x_2)
+\Delta\bar q_i(x_1)\Delta q_i(x_2)]}
{\sum_ie_i^2[ q_i(x_1)\bar q_i(x_2)
+\bar q_i(x_1) q_i(x_2)]}
\end{equation}
and
\begin{equation}
A_{TT}^{l\bar l}=a_{TT}\frac{\sum_ie_i^2[\delta q_i(x_1)\delta\bar q(x_2)
+\delta\bar q_i(x_1)\delta q_i(x_2)]}
{\sum_ie_i^2[ q_i(x_1)\bar q_i(x_2)
+\bar q_i(x_1) q_i(x_2)]},
\end{equation}
where $x_1x_2=M^2_{l\bar l}/s$ and $x_F=x_1-x_2$.

The ratio of the asymmetries
$A_{TT}^{l\bar l}$ and $A_{LL}^{l\bar l}$ is also small in the central
region of low-mass Drell-Yan production:
\begin{equation}
A_{TT}^{l\bar l}/A_{LL}^{l\bar l}\simeq 0
\end{equation}
This result agrees with the predictions
made in \cite{jafs}.

The above results were obtained in the limit $s\to \infty$ which
corresponds to the limit $x\to 0$, i.e. they have an asymptotic
nature.
However, it might happen that the kinematical region of the SMC
experiment lies in the preasymptotic domain and the above formulas
in fact
are valid at much smaller values of $x$, than the range covered
by the present experiments.

Indeed, it was shown that the
preasymptotic effects are very important to understand
the experimental regularities
observed in the unpolarized scattering \cite{pras}.

The direct calculation
of the quark densities at small $x$ in the preasymptotic region $x>x_0$
according to Eqs. (\ref{h1},\ref{f2}) and (\ref{u13}) provides
the following expressions:
\begin{equation}
 q(x)\simeq \frac{c_1}{x}(1+\frac{c_2}{\sqrt{x}}),\label{pf1}
\end{equation}
\begin{equation}
\Delta q(x)\simeq \frac{c_3}{x}(1+\frac{c_4}{\sqrt{x}})\label{pg1}
\end{equation}
and
\begin{equation}
\delta q(x)\simeq {c_5}(1+\frac{c_6}{\sqrt{x}}),\label{ph1}
\end{equation}
where the parameters $c_{1,2}\geq 0$.  The parameters
$c_{i}$, $(i=3-6)$  can get negative as well as
positive values.
It means that in the preasymptotic region the structure
 functions $g_1$ and $h_1$ may demonstrate a sign changing behaviour.
Using the above expressions we arrive again to small values of transverse
double--spin asymmetries
$A_{TT}^{l\bar l}$  in the central
region of low-mass Drell-Yan production, but the corresponding
longitudinal asymmetries would  not disappear in the preasymptotic
region.
Thus, the measurements of the double--spin asymmetries at
 RHIC
would provide important information on $x$-dependence of the spin
quark densities $\Delta q(x)$ and $\delta q(x)$.

\section*{Conclusion}

We  considered the low-$x$ behaviour of the spin structure functions
$g_1(x)$ and $h_1(x)$ in the framework of the nonperturbative approach
on partonic structure
 of the constituent quarks and their scattering
picture in the  effective two-component field.  We
 obtained
 the spin densities of current quarks.  Such considerations
could be regarded as a kind of a bootstrap approach.

The obtained  singular dependence of $g_1(x)$  at $x\to 0$
can describe  the rise of the spin structure function observed
in the SMC experiment. The model predicts smooth behaviour of $h_1(x)$
at $x\to 0$ which, contrary to $g_1$, corresponds to the leading diffractive contribution.

It was shown that the
 singular behaviour of the structure function $g_1(x)$ does not affect
much the numerical results on the partition of a nucleon spin.

Vanishingly small
 double--spin asymmetries in the Drell-Yan production with low
invariant masses were predicted in the central region.

\newpage
\section*{Figure captions}
\bf Fig. 1
\rm The $x$--dependence of the proton spin structure function
$g_1^p(x)$.\\ [2ex]
\bf Fig. 2
\rm The $x$--dependence of the neutron spin structure function
$g_1^n(x)$.
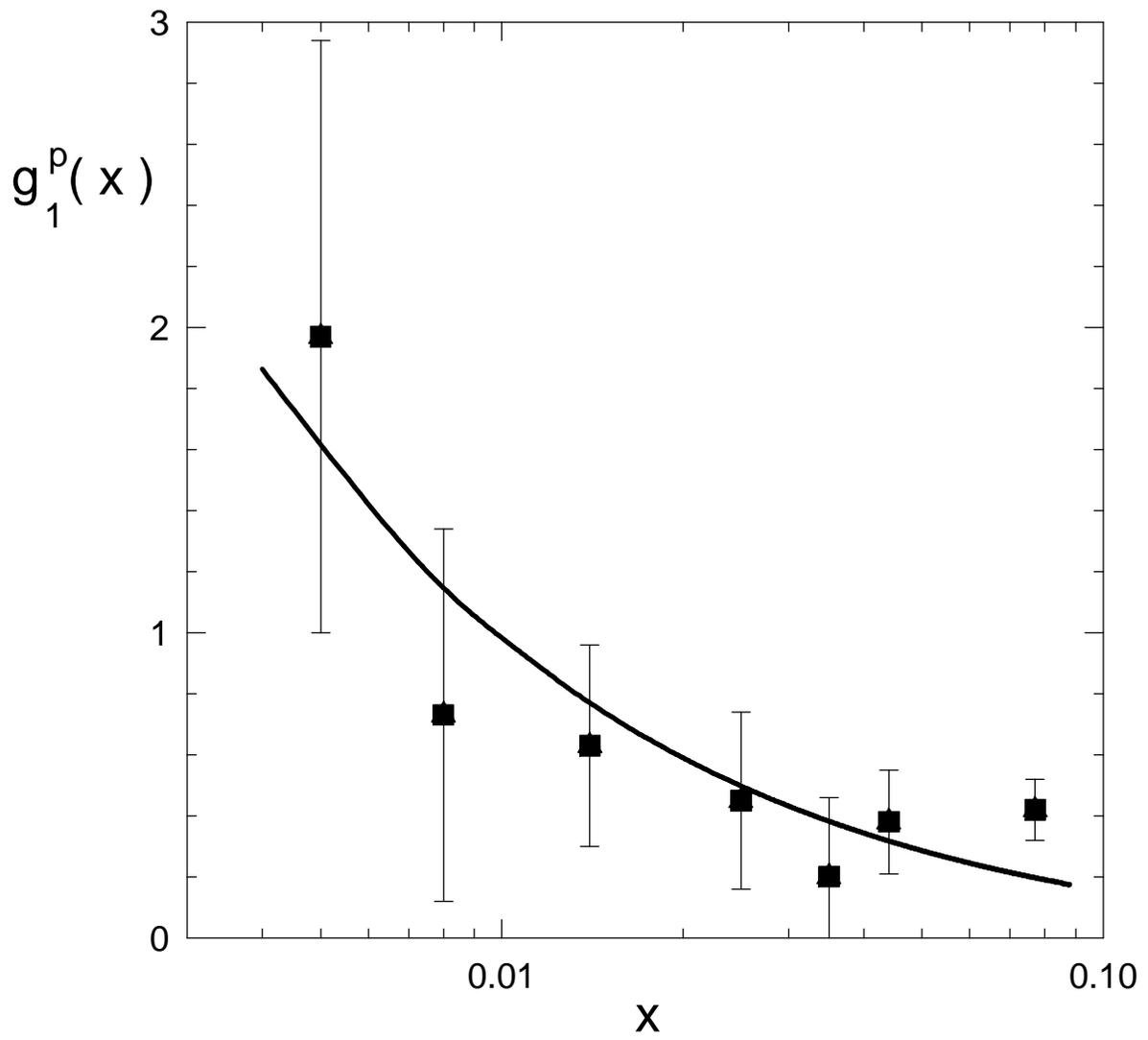
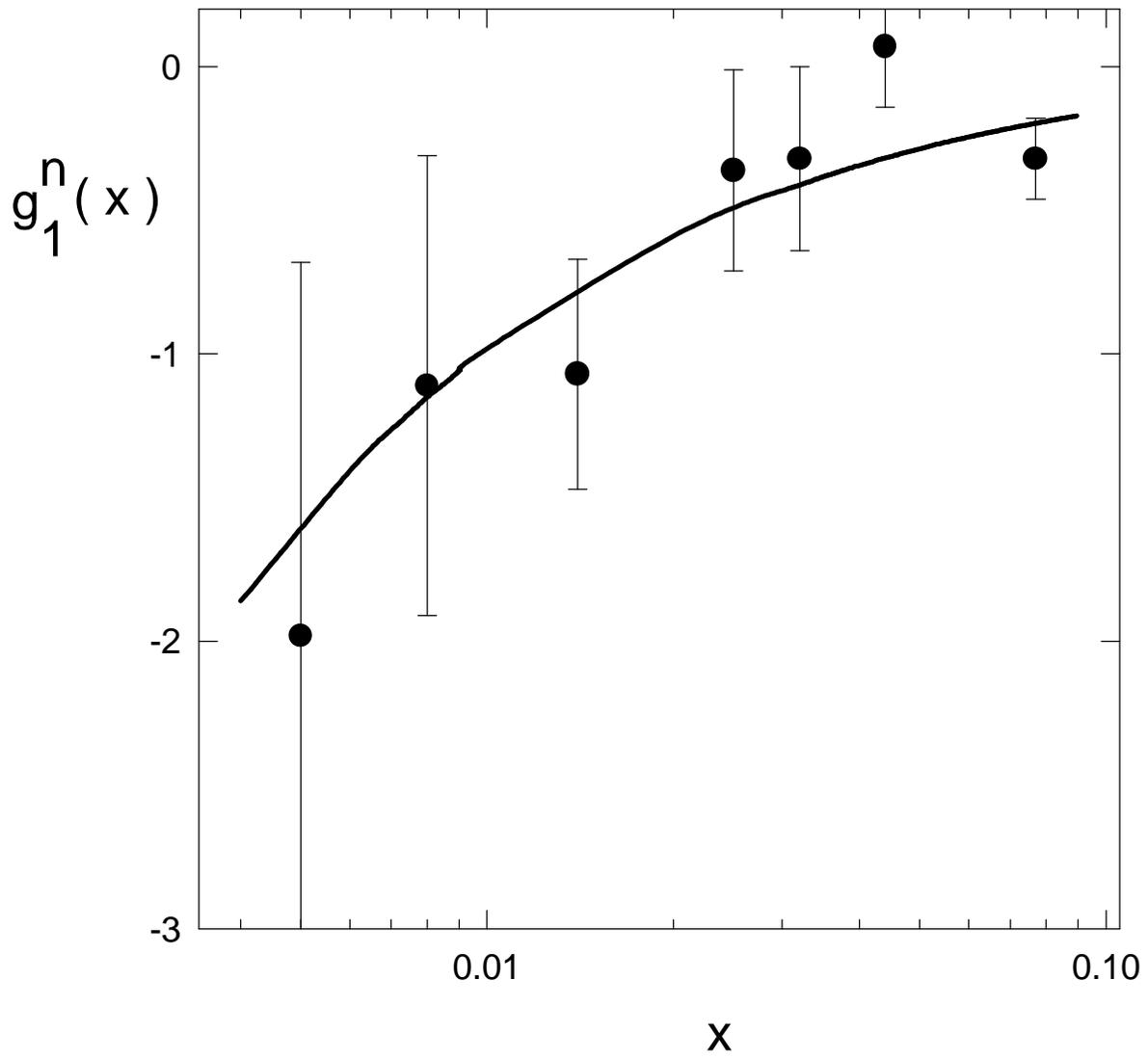
\end{document}